\documentclass{article}
\usepackage[a4paper, left=1.5cm, right=1.5cm, top=2.5cm, bottom=2.5cm]{geometry}
\usepackage{graphicx} % Required for inserting images
\usepackage{amsmath}
\usepackage{amssymb}
\usepackage{xcolor}
\usepackage{authblk}
\usepackage{cite}
\usepackage{subcaption}
\usepackage{float}
\usepackage{placeins}
\usepackage{comment}
\title{Constraining Cosmic Deceleration using the Solar System}
\author{Paridhi Jain $^{1}$ \thanks{paridhi.j@ahduni.edu.in}, Oem Trivedi $^{2}$ \thanks{oem.trivedi@vanderbilt.edu}, Maxim Khlopov $^{1}$ \thanks{maxim.khlopov@cern.ch},Jackson Levi Said $^{3,4}$ \thanks{jackson.said@um.edu.mt}, Maxim Krasnov $^{5,6}$ \thanks{morrowindman1@mail.ru}}
\affil[1]{Virtual Institute of Astroparticle Physics, Paris 75018, France}
\affil[2]{Department of Physics and Astronomy, Vanderbilt University, Nashville, TN 37235, USA}
\affil[3]{Institute of Space Sciences and Astronomy, University of Malta, Msida MSD 2080, Malta}
\affil[4]{Department of Physics, University of Malta, Msida, Malta}
\affil[5]{Research Institute of Physics, Southern Federal University, Rostov-on-Don 344090, Russia}
\affil[6]{National Research Nuclear University MEPhI, Moscow 115409, Russia}
\begin{document}

\maketitle

\begin{abstract}
We investigate whether Solar System dynamics can provide independent constraints on the present day cosmological deceleration parameter $q_0$, offering a complementary approach to conventional cosmological probes. By relating cosmic acceleration to a weak Hooke type radial perturbation, we derive bounds using planetary perihelion precession, trans-Neptunian objects, and Oort cloud dynamics. We obtain relatively robust planetary bounds of $|q_0|\lesssim 10^9$, while the outer Solar System yields potentially stronger, albeit more model dependent, constraints approaching $10^5$. Through Monte Carlo uncertainty propagation and sensitivity analyses, we explore the statistical robustness of these estimates and their dependence on orbital measurements. Finally, we demonstrate how future precision astrometry and outer planet missions could improve these limits by several orders of magnitude, establishing Solar System dynamics as a complementary laboratory for probing cosmological acceleration.
\end{abstract}
\section{Introduction}
Modern cosmology is presently undergoing a period of remarkable theoretical and observational development, with advances in both domains allowing us to investigate the evolution of the Universe across an increasingly broad range of physical scales and cosmic epochs. One of the most significant developments in this regard has been the extensive exploration of dark energy models \cite{de1SupernovaSearchTeam:1998fmf,de2Li:2012dt,de3Li:2011sd,de4Mortonson:2013zfa,de5Frusciante:2019xia,de6Huterer:2017buf,de7Vagnozzi:2021quy,de8Adil:2023ara,de9Feleppa:2025clx,de10DiValentino:2020evt,de11Nojiri:2010wj,de12Nojiri:2006ri,de13Trivedi:2023zlf,de14Trivedi:2022svr,de15Trivedi:2024inb,de16Trivedi:2024dju}, which have considerably broadened our understanding of the possible mechanisms governing cosmic acceleration and the late time expansion history of the Universe. Despite the remarkable success of the standard $\Lambda$CDM model, persistent observational discrepancies, particularly the $H_0$ and $S_8$ tensions \cite{ht1DiValentino:2021izs,ht2Clifton:2024mdy,s81kazantzidis2018evolution,s82amon2022non,s83poulin2023sigma,baoFerreira:2025lrd}, continue to motivate investigations into physics beyond the conventional cosmological framework. At the same time, the rapidly expanding landscape of observational cosmology including the increasingly precise measurements of the cosmic microwave background, large scale structure, high redshift astrophysical populations, gravitational waves and the cosmic web, promises to provide unprecedented opportunities for testing competing theoretical descriptions of the Universe \cite{t1jha2019next,t2Shanks:2015lda,t3chandler2025nsf,t4Euclid:2024yrr,t5WST:2024rai,t6CosmoVerseNetwork:2025alb,t7COMPACT:2022gbl}. The interplay between these developments has consequently placed some of the most fundamental questions concerning gravity, spacetime and cosmic evolution at the forefront of contemporary research.
\\
\\
Within this increasingly diverse theoretical landscape, scalar field descriptions of dark energy have attracted particular attention owing to their flexibility in accommodating different cosmic expansion histories. These include canonical quintessence, non-minimally coupled scalar tensor theories, k-essence, Horndeski constructions, modified gravity inspired scenarios and models involving non-canonical kinetic interactions \cite{de2Li:2012dt,de3Li:2011sd}. However, the observational foundations of fundamental physics are not restricted exclusively to measurements on cosmological scales. Solar System observations and other local experiments have historically provided stringent tests of gravitational theories, particle physics and possible interactions involving the dark sector. In particular, investigations of dark matter have benefited considerably from direct detection experiments, precision measurements of orbital dynamics and complementary astrophysical observations \cite{dm11rubin1970rotation,dm1Cirelli:2024ssz,dm2Arbey:2021gdg,dm3Balazs:2024uyj,dm4Eberhardt:2025caq,dm5Bozorgnia:2024pwk,dm6Misiaszek:2023sxe,dm7OHare:2024nmr,dm8Adhikari:2022sbh,dm9Miller:2025yyx,dm10Trivedi:2025vry}. Similarly, pulsar timing observations have emerged as valuable probes of ultralight fields and compact astrophysical objects, while precision orbital measurements provide important restrictions on possible deviations from established gravitational dynamics, including additional long range interactions \cite{pta1Gan:2025icr,pta2Unal:2022ooa,pta3Ramani:2020hdo}. Dark energy, however, has received comparatively limited attention in this local context, primarily because of its extremely small characteristic energy density and its association with the expansion of the Universe on cosmological scales. Nevertheless, certain dynamical dark energy and modified gravity scenarios can, depending on their interactions and screening mechanisms, introduce additional contributions to local gravitational dynamics or variations in fundamental quantities. This raises the interesting possibility that precision observations within gravitationally bound systems might provide complementary information about the physical mechanisms underlying late time cosmic evolution.
\\
\\
Recently, an interest has spurred towards the notion of cosmic deceleration and its possibilities with regards to new surveys, particularly as increasingly precise cosmological measurements and reanalyses of existing observations have renewed discussions concerning the possibility of an evolving expansion history \cite{dec1sah2026pantheon+,dec3tangphati2026dark,dec4nielsen2016marginal,dec5sah2025anisotropy,dec6wang2025questioning,dec7ferri2026present}. While the standard cosmological model describes a Universe currently undergoing accelerated expansion, departures from a strictly constant dark energy component could potentially alter the future evolution of cosmic acceleration, including the possibility of a transition towards deceleration in certain theoretical scenarios. Understanding the extent to which such possibilities are observationally distinguishable is hence important, not only for evaluating the robustness of the standard cosmological picture but also for examining the broader range of expansion histories that remain compatible with observations. The cosmological deceleration parameter, $q=-\ddot{a}/(aH^2)$, provides a particularly useful quantity in this context, as it directly characterizes the acceleration or deceleration of the cosmic scale factor. Conventionally, its determination relies upon cosmological measurements but it is worthwhile to see whether complementary information can also be obtained from the dynamics of gravitationally bound systems. In particular, a cosmological contribution to the local radial acceleration can, under appropriate assumptions, be expressed through a weak Hooke type perturbation proportional to the orbital radius, which can end up providing us with a possible connection between precision orbital measurements and the cosmological expansion history. This motivates us to investigate whether anomalous accelerations, orbital precession measurements and the dynamics of distant Solar System populations can place meaningful restrictions on the present day deceleration parameter. Hence, we try to consider the possibility of constraining cosmic deceleration using the variety of probes we can get at local solar system level. We note that these constraints are not supposed to be competitive to CMB or Galaxy level probes, however, and instead we are concerned with figuring out the limits of solar system constraints on relevant cosmological quantities. 
\\

\section{Cosmological Constraints from the Solar System}
The present day cosmological deceleration parameter provides a direct characterization of the acceleration or deceleration of the cosmic expansion and is hence an important quantity in understanding the late time evolution of the Universe. Within the standard $\Lambda$CDM framework, the present epoch is characterized by accelerated expansion, corresponding to $q_0<0$, with the observational determination of this quantity relying predominantly on cosmological measurements such as Type Ia supernovae, the cosmic microwave background and baryon acoustic oscillations. However, recent developments have motivated renewed investigations into the possibility that the cosmic expansion history may depart from the conventional picture. In particular, Sah, Rameez and Sarkar \cite{sah2026pantheon+} have presented a reanalysis of the Pantheon+ supernova compilation incorporating corrections associated with progenitor age, finding results that challenge the conventional interpretation of late-time cosmic acceleration. Although the implications of such analyses remains under debate, they highlight the importance of understanding the observational assumptions underlying our reconstruction of the expansion history. Simultaneously, recent results from the Dark Energy Spectroscopic Instrument (DESI), particularly when combined with supernova and CMB measurements, have indicated a possible preference for an evolving dark energy equation of state over a strictly constant cosmological constant \cite{adame2025desi,zhang2026}. Such developments raise important questions regarding the evolution of the deceleration parameter, the possibility of transitions between accelerating and decelerating phases, and the observational methods through which these possibilities could be independently examined.
\\ 
\\
An interesting direction in this regard is to explore whether gravitational dynamics within the Solar System can provide complementary information about cosmological quantities. The Solar System has historically served as one of the most precise laboratories for testing gravitational physics, with measurements of planetary positions, spacecraft trajectories and orbital precession providing stringent restrictions on possible deviations from Newtonian gravity and general relativity. Modern planetary ephemerides, including the EPM series, incorporate increasingly accurate observational information and detailed modeling of gravitational perturbations, making it possible to investigate small anomalous contributions to the orbital dynamics of planets and spacecraft \cite{pitjeva2013relativistic,pitjeva2013updated,pitjev2013constraints}. Furthermore, investigations of the influence of a cosmological constant on the gravitational two-body problem have demonstrated how precision orbital observables can be employed to constrain additional radial accelerations and related departures from conventional gravitational dynamics \cite{jetzer2006two,sereno2006solar,iorio2008solar}. Such studies provide an important precedent for considering whether quantities traditionally associated with cosmological evolution might leave detectable, albeit extremely suppressed, signatures in local gravitational systems. The particular advantage of Solar System observations is that they probe gravitational dynamics in a regime vastly different from that of cosmological surveys, potentially offering an independent means of testing the consistency of gravitational descriptions across different physical scales. At the same time, the interpretation of any anomalous orbital contribution must carefully account for ordinary gravitational perturbations, parameter degeneracies and the possibility that gravitationally bound systems are effectively decoupled from aspects of the cosmological expansion \cite{wang2023decoupling}.
\\
\\
Beyond the planetary region, the outer Solar System presents additional possibilities for investigating extremely weak gravitational effects. The discovery and subsequent characterization of distant trans-Neptunian objects, including Sedna, have opened new avenues for examining orbital dynamics at distances considerably larger than those accessible through the classical planetary system \cite{brown2004discovery}. At still greater distances, the Oort cloud represents an interesting dynamical environment in which solar gravitational attraction competes with perturbations arising from the Galactic tidal field, passing stars and the long-term evolution of the Solar System itself. Previous investigations have explored the formation and dynamical evolution of the Oort cloud, the influence of Galactic tides and the possible consequences of additional non-Newtonian accelerations on the motion of distant Solar System bodies \cite{brasser2013oort,klacka2010galactic,heisler,torres2019galactic,iorio2012impact}. The importance of these distant systems for our present investigation follows from the fact that certain weak radial perturbations increase in magnitude with distance, even when their effects are negligibly small within the inner planetary region. Consequently, outer Solar System objects may provide interesting complementary sensitivity to such perturbations, although their observational baselines, complex dynamical histories and susceptibility to external gravitational influences introduce substantial uncertainties. It is hence very useful to investigate planetary ephemerides, trans-Neptunian objects, long period comets and the Oort cloud within a common theoretical framework, while recognizing that the observational reliability and model dependence of the corresponding estimates can differ significantly.
\\
\\
With these considerations in mind, we investigate a phenomenological connection between the cosmological expansion history and a weak additional radial acceleration in the Solar System. In an appropriate local description of relative motion embedded within an expanding cosmological background, the contribution associated with the background expansion can be represented, under suitable approximations, by a Hooke-type term proportional to the separation between gravitating bodies. We begin by defining the present day cosmological acceleration coefficient
\begin{equation}
K_0 \equiv \frac{\ddot a}{a}
\end{equation}
and use the standard relation between this quantity and the present day deceleration parameter
\begin{equation}
K_0 = -q_0H_0^2
\end{equation}
where $H_0$ denotes the present day Hubble parameter. Within the adopted effective description, the corresponding radial acceleration takes the form
\begin{equation}
A_K = K_0r = -q_0H_0^2r
\end{equation}
This provides the central mathematical correspondence employed throughout our investigation. In particular, an observational restriction on an additional acceleration proportional to the orbital radius can be translated into a restriction on the effective coefficient $K_0$, which may subsequently be related to $q_0$ through the cosmological relation given above. It is important to emphasize that this identification assumes that the relevant local perturbation can be consistently associated with the cosmological background acceleration as in general, the dynamics of gravitationally bound systems need not respond to the global expansion in a universal manner, and additional contributions arising from local matter distributions, modified gravity or other physical mechanisms may exhibit similar radial dependencies. Accordingly, the resulting translation to $q_0$ should be understood within the assumptions of the adopted effective description rather than as an entirely model-independent correspondence.
\\
\\
For our numerical estimates, we adopt the reference value
\begin{equation}
H_0 = 67.4\,{\rm km\,s^{-1}\,Mpc^{-1}}
\simeq 6.89 \times 10^{-11}\,{\rm yr}^{-1}
\end{equation}
which gives
\begin{equation}
H_0^2 \simeq 4.75 \times 10^{-21}\,{\rm yr}^{-2}
\end{equation}
Note that the two representations of $H_0$ are equivalent, since $1\,{\rm Mpc} \simeq 3.08568 \times 10^{19}\,{\rm km}$ and $1\,{\rm yr} \simeq 3.15576 \times 10^7\,{\rm s}$. While the conventional units of ${\rm km\,s^{-1}\,Mpc^{-1}}$ facilitate direct comparison with cosmological measurements, we employ inverse year units in our calculations for consistency with the astronomical timescales used in orbital dynamics and secular precession analyses. In particular, expressing $H_0^2$ in ${\rm yr}^{-2}$ will allow its contribution to the effective radial acceleration to be evaluated directly alongside the corresponding orbital quantities. The extremely small magnitude of $H_0^2$ in local dynamical units has an immediate consequence for the observational prospects of this approach. Even highly restrictive limits on an additional radial acceleration within the Solar System can correspond to comparatively weak limits on the dimensionless cosmological deceleration parameter. This tells us why local orbital measurements are not expected to compete directly with cosmological observations in determining a deceleration parameter of order unity. Still, the exercise remains valuable for establishing the extent to which precision gravitational measurements can probe cosmologically motivated perturbations, identifying which observational systems offer the greatest sensitivity and clarifying the assumptions under which local dynamical constraints may be connected to large-scale cosmological evolution. With this motivation, we shall proceed to investigate the constraints obtainable from different Solar System observables, beginning with planetary orbital dynamics and subsequently extending the analysis to more distant gravitationally bound populations.
\\

\section{Deceleration Parameter constraints} 
The framework developed in the preceding section allows us to investigate several complementary Solar System observables as potential probes of the cosmological deceleration parameter. In particular, we will consider planetary perihelion precession measurements, constraints associated with distributed matter and cosmological contributions to local dynamics, and the orbital properties of more distant Solar System populations. These approaches offer different levels of observational precision and theoretical sensitivity, making it important to distinguish direct ephemeris-based estimates from approximate dynamical consistency arguments. With this in mind, we now examine the constraints obtainable through each of these routes and assess their respective implications for the possibility of probing cosmic acceleration or deceleration through local gravitational measurements.
\\

Before we get towards the constraints, we should discuss briefly the screening mechanisms and matter couplings associated with dark scalars which would make the constraints below relevant. In general, the presence of a cosmologically evolving scalar field does not necessarily imply that its effects can be detected through the dynamics of gravitationally bound systems. For the local constraints derived in this work to be physically meaningful, we require a theory in which the cosmological background can communicate with local gravitational dynamics, either through the metric or through an additional scalar mediated interaction. A useful starting point is a scalar-tensor action written in the Einstein frame as
\begin{equation}
S = \int d^4x \sqrt{-g}\left[\frac{M_{\rm Pl}^2}{2}R-\frac{1}{2}g^{\mu\nu}\partial_\mu\phi\partial_\nu\phi-V(\phi)\right]+S_m[\psi_m,A^2(\phi)g_{\mu\nu}]
\end{equation}
where $M_{\rm Pl}$ is the reduced Planck mass, $V(\phi)$ is the scalar potential and $A(\phi)$ determines the coupling between the scalar and ordinary matter. The corresponding dimensionless coupling strength can be expressed as
\begin{equation}
\beta(\phi)=M_{\rm Pl}\frac{d\ln A(\phi)}{d\phi}
\end{equation}
For nonrelativistic matter, the scalar contribution to the local acceleration is then approximately $\mathbf{a}_{\phi}=-\frac{\beta(\phi)}{M_{\rm Pl}}\boldsymbol{\nabla}\phi$. This tells us that the presence of a matter coupling alone is not sufficient for obtaining the radial acceleration considered in our analysis. The scalar configuration must also possess an appropriate spatial response and its contribution must remain sufficiently smooth over the orbital region being investigated. In particular, a generic scalar fifth force need not have the same radial dependence as the cosmological term $A_K=K_0r$, so the two contributions should not automatically be identified. The screening of such scalar interactions becomes important because local gravitational experiments already impose strong restrictions on additional long range forces. For example, in chameleon type theories the scalar responds to the surrounding matter density through an effective potential of the form
\begin{equation}
V_{\rm eff}(\phi,\rho_m)=V(\phi)+\rho_m A(\phi)
\end{equation}
where $\rho_m$ denotes the conserved nonrelativistic matter density in the adopted Einstein frame description, the effective scalar mass is determined by
\begin{equation}
m_{\rm eff}^2(\rho_m)=\left.\frac{\partial^2 V_{\rm eff}}{\partial\phi^2}\right|_{\phi=\phi_{\rm min}(\rho_m)}
\end{equation}
and a sufficiently large value of $m_{\rm eff}$ in a dense environment suppresses the propagation of the scalar interaction. Another possibility is symmetron screening, for which one may consider
\begin{equation}
V(\phi)=-\frac{1}{2}\mu^2\phi^2+\frac{\lambda}{4}\phi^4,\qquad A(\phi)\simeq 1+\frac{\phi^2}{2M^2}
\end{equation}
In this case the effective coupling becomes small when the environmental density restores the symmetry and drives $\phi$ towards zero. Derivative screening, including the Vainshtein mechanism, offers a different possibility in which nonlinear kinetic interactions suppress scalar gradients in the vicinity of massive bodies. These mechanisms are relevant for our investigation because the Solar System can lie in a screened environment even when a scalar field contributes appreciably to cosmological evolution. Consequently, the existence of a dynamical dark energy component does not by itself guarantee that its local effects can be constrained through planetary orbits.
\\
\\
To understand the requirements more explicitly, we can write the effective radial acceleration experienced by a test body as
\begin{equation}
A_r(r)=-\frac{GM_\odot}{r^2}+K_{\rm bg}(t)r+A_\phi(r,\rho_m)
\end{equation}
where $K_{\rm bg}(t)$ describes the cosmological background contribution in the adopted local metric, while $A_\phi$ represents the additional scalar mediated force. The identification used throughout this work requires $K_{\rm bg}(t_0)=-q_0H_0^2$ and assumes that the remaining scalar contribution can either be neglected or consistently included in an effective Hooke like coefficient. For a sufficiently smooth scalar configuration, we may expand its acceleration around a reference orbital radius $r_*$ as
\begin{equation}
A_\phi(r,\rho_m)=A_\phi(r_*,\rho_m)+\left.\frac{\partial A_\phi}{\partial r}\right|_{r_*}(r-r_*)+\mathcal{O}((r-r_*)^2)
\end{equation}
However, this expansion does not generally produce a universal acceleration proportional to $r$. A useful phenomenological requirement is therefore that the observable additional acceleration can be approximated over the relevant orbital domain by
\begin{equation}
A_{\rm extra}(r)\simeq K_{\rm eff}r,\qquad K_{\rm eff}=-q_0H_0^2+\Delta K_\phi
\end{equation}
where $\Delta K_\phi$ denotes any scalar contribution having the same effective radial dependence. A direct translation of an orbital bound into a constraint on $q_0$ is justified only when $\Delta K_\phi$ is negligible, independently constrained or related to the background evolution through a specified scalar-tensor theory. Otherwise, the observational bound should be interpreted as a constraint on $K_{\rm eff}$ rather than on the cosmological deceleration parameter alone. An additional requirement is that the screening response should be compatible with the different gravitational environments considered below. For instance, a scalar may be strongly screened near the Sun and the inner planets while its effective mass or coupling may change in the lower density regions occupied by distant trans Neptunian objects and the Oort cloud. In such a case, the scalar contribution would be environmentally dependent, and the same constant $K_0$ need not describe all orbital populations. Schematically, the relevant condition for interpreting our bounds through a common cosmological coefficient is
\begin{equation}
\left|A_\phi(r,\rho_m)-\Delta K_\phi r\right|\ll |K_{\rm eff}r|
\end{equation}
over the radial range under consideration, together with a sufficiently slow temporal evolution of the effective coefficient over the observational baseline. If the scalar is efficiently screened and the cosmological background term survives in the local dynamics, the constraints can be interpreted through $K_0=-q_0H_0^2$ as assumed in our framework. If both the scalar interaction and the background response are suppressed, the absence of anomalous orbital effects does not necessarily constrain $q_0$. On the other hand, if an unscreened scalar generates an additional long range force, a model-specific treatment is required to distinguish its contribution from the cosmological acceleration. Hence, the bounds obtained in the following subsections should be understood as constraints on the effective Hooke-like acceleration, with their interpretation in terms of cosmic deceleration depending on the underlying matter coupling, screening efficiency and local response to the cosmological background.

\subsection{Limits from Saturn and Iorio type bounds}
Iorio considered the general case in which the relative acceleration rate of the scale factor is slowly time dependent and this amounts to expanding
\begin{equation}
K(t) = \frac{\ddot a}{a} = K_0 + K_1(t-t_0)
\end{equation}
with the corresponding perturbing acceleration
\begin{equation}
A(t) = K(t)r
\end{equation}
The first term $K_0r$ produces the usual Hooke-like perturbation, while the term $K_1(t-t_0)r$ captures the slow local time variation of cosmic acceleration. Iorio's quoted Saturn bound is most directly a constraint on $K_1$, not on $K_0$. The Saturn based constraint obtained from the anomalous perihelion precession is
\begin{equation}
K_1 \lesssim 2 \times 10^{-13} {\rm yr}^{-3}
\end{equation}
This is a limit on $d(\ddot a/a)/dt$ and it is hence directly useful for rapidly evolving dark energy models or imminent singularity models, but it is not by itself a model-independent bound on $q_0$. To see this explicitly, we differentiate
\begin{equation}
K = -qH^2
\end{equation}
Using
\begin{equation}
\dot H = -(1+q)H^2
\end{equation}
one obtains
\begin{equation}
\dot K = -\dot q H^2 + 2q(1+q)H^3
\end{equation}
Therefore, the Saturn bound on $K_1$ gives
\begin{equation}
\left|-\dot q_0H_0^2 + 2q_0(1+q_0)H_0^3\right| \lesssim 2 \times 10^{-13} {\rm yr}^{-3}
\end{equation}
Dividing by $H_0^3$ gives
\begin{equation}
\left|-\frac{\dot q_0}{H_0} + 2q_0(1+q_0)\right| \lesssim \frac{2 \times 10^{-13}}{H_0^3}
\end{equation}
With the numerical value of $H_0$ adopted above, it is easy to check that
\begin{equation}
H_0^3 = 3.27 \times 10^{-31} {\rm yr}^{-3}
\end{equation}
and hence
\begin{equation}
\left|-\frac{\dot q_0}{H_0} + 2q_0(1+q_0)\right| \lesssim 6.1 \times 10^{17}
\end{equation}
This is the most faithful way of translating Iorio's $K_1$ constraint and let's see what it really tells us. It shows that the actual observable is a combination of $q_0$ and $\dot q_0$ and without an assumption about the local time evolution of $q$, this cannot be converted into a clean bound on $q_0$ alone. If one imposes the additional assumption that $q$ is approximately constant over the relevant solar system timescale, then
\begin{equation}
\dot q_0 \simeq 0
\end{equation}
and the previous bound reduces to
\begin{equation}
|2q_0(1+q_0)| \lesssim 6.1 \times 10^{17}
\end{equation}
For $|q_0| \gg 1$, this becomes
\begin{equation}
2q_0^2 \lesssim 6.1 \times 10^{17}
\end{equation}
so that
\begin{equation}
|q_0| \lesssim 5.5 \times 10^8
\end{equation}
This is a useful approximate number, but it is conditional on the assumption that $\dot q_0$ is negligible and so, it should not be interpreted as a fully model-independent solar system bound on the deceleration parameter. A cleaner way of extraction of a constant $K_0$ constraint can be obtained by using the same Saturn perihelion residual, but applying it to the constant Hooke-like acceleration $A=K_0r$. For such a perturbation, the secular perihelion precession is \cite{jetzer2006two,sereno2006solar,iorio2008solar}
\begin{equation}
\langle \dot\varpi \rangle = \frac{3K_0}{2n}\sqrt{1-e^2}
\end{equation}
where $n=2\pi/P$ is the Keplerian mean motion and solving for $K_0$ gives
\begin{equation}
|K_0| \lesssim \frac{2n|\Delta \dot\varpi|}{3\sqrt{1-e^2}}
\end{equation}
For Saturn we use \cite{pitjeva2013updated}
\begin{equation}
P_{\rm Sat} = 29.46 {\rm yr}
\end{equation}
\begin{equation}
n_{\rm Sat} = \frac{2\pi}{P_{\rm Sat}} = 0.213 {\rm yr}^{-1}
\end{equation}
\begin{equation}
e_{\rm Sat} = 0.056
\end{equation}
and the perihelion residual \cite{pitjeva2013relativistic,pitjeva2013updated}
\begin{equation}
\Delta \dot\varpi_{\rm Sat} \lesssim 0.47 {\rm mas \ cty}^{-1}
\end{equation}
Using $1 {\rm mas} = 4.848 \times 10^{-9} {\rm rad}$ one obtains $0.47 {\rm mas \ cty}^{-1} = 2.28 \times 10^{-11} {\rm rad \ yr}^{-1}$ and this substitution then gives us
\begin{equation}
|K_0|_{\rm Sat} \lesssim 3.2 \times 10^{-12} {\rm yr}^{-2}
\end{equation}
and hence
\begin{equation}
|q_0|_{\rm Sat} \lesssim \frac{3.2 \times 10^{-12}}{4.75 \times 10^{-21}}
\end{equation}
Therefore
\begin{equation}
|q_0|_{\rm Sat} \lesssim 6.8 \times 10^8
\end{equation}
This is the most direct Saturn based estimate for a constant deceleration parameter. It is numerically close to the value inferred from the $K_1$ bound under the slow $q$ assumption, but conceptually cleaner because it constrains $K_0$ rather than $\dot K$. \\

\subsection{Limits from the Mars route}
Mars provides us with a complementary route because its ranging data are very accurate. \cite{pitjeva2013updated,fienga2011,kuchynka} Although the cosmological Hooke like precession is less enhanced for Mars than for Saturn due to Mars having a larger mean motion, the quality of ranging data compensates for this to some extent. The same constant $K_0$ perihelion formula gives
\begin{equation}
|K_0|_{\rm Mars} \lesssim \frac{2n_{\rm Mars}|\Delta \dot\varpi_{\rm Mars}|}{3\sqrt{1-e_{\rm Mars}^2}}
\end{equation}
For Mars we take \cite{pitjeva2013updated}
\begin{equation}
P_{\rm Mars} = 1.8808 {\rm yr}
\end{equation}
\begin{equation}
n_{\rm Mars} = \frac{2\pi}{P_{\rm Mars}} = 3.34 {\rm yr}^{-1}
\end{equation}
\begin{equation}
e_{\rm Mars} = 0.0935
\end{equation}
A representative anomalous perihelion sensitivity can be taken as \cite{pitjeva2013relativistic,pitjeva2013updated}
\begin{equation}
\Delta \dot\varpi_{\rm Mars} \sim 0.037 {\rm mas \ cty}^{-1}
\end{equation}
This corresponds to
\begin{equation}
\Delta \dot\varpi_{\rm Mars} = 0.037 \times \frac{4.848 \times 10^{-9}}{100} = 1.79 \times 10^{-12} {\rm rad \ yr}^{-1}
\end{equation}
Substitution of this gives
\begin{equation}
|K_0|_{\rm Mars} \lesssim 4.0 \times 10^{-12} {\rm yr}^{-2}
\end{equation}
and therefore
\begin{equation}
|q_0|_{\rm Mars} \lesssim \frac{4.0 \times 10^{-12}}{4.75 \times 10^{-21}} \lesssim 8.5 \times 10^8
\end{equation}
The Mars perihelion route hence gives us a bound very close to the Saturn based estimate. This shows that while Mars ranging is already competitive with Saturn style limits, but it does not improve the constraint by orders of magnitude. Now, if the allowed anomalous perihelion interval is taken more broadly as
\begin{equation}
\Delta \dot\varpi_{\rm Mars} \sim 0.02{\rm \ to \ }0.11 {\rm mas \ cty}^{-1}
\end{equation}
then the same scaling gives
\begin{equation}
|q_0|_{\rm Mars} \sim 5 \times 10^8{\rm \ to \ }3 \times 10^9
\end{equation}
This range should be interpreted as an order of magnitude sensitivity estimate rather than a dedicated ephemeris fit. The clean way to obtain the final value would be to include $K_0$ directly as a fitted parameter in a modern planetary ephemeris. 
\\
\\
There is also a useful Mars based estimate from the local dark matter density constraint, as a uniform density produces a Hooke-like acceleration \cite{local,iorio2006solar}
\begin{equation}
A_\rho = -\frac{4\pi G\rho}{3}r
\end{equation}
so that the corresponding effective Hooke coefficient is
\begin{equation}
|K_\rho| = \frac{4\pi G\rho}{3}
\end{equation}
Using the representative Mars distance bound $\rho_{\rm dm} \lesssim 1.4 \times 10^{-20} {\rm g \ cm}^{-3}$ \cite{pitjev2013constraints,pitjeva2013relativistic} and converting units gives $1.4 \times 10^{-20} {\rm g \ cm}^{-3} = 1.4 \times 10^{-17} {\rm kg \ m}^{-3}$. So from this we get
\begin{equation}
|K_\rho| \lesssim \frac{4\pi}{3}(6.674 \times 10^{-11})(1.4 \times 10^{-17}) {\rm s}^{-2}
\end{equation}
It is easy to check that
\begin{equation}
|K_\rho| \lesssim 3.9 \times 10^{-27} {\rm s}^{-2}
\end{equation}
Using $1 {\rm yr} = 3.15576 \times 10^7 {\rm s}$ this becomes
\begin{equation}
|K_\rho| \lesssim 3.9 \times 10^{-12} {\rm yr}^{-2}
\end{equation}
and so we have
\begin{equation}
|q_0| \lesssim \frac{3.9 \times 10^{-12}}{4.75 \times 10^{-21}} \lesssim 8.2 \times 10^8
\end{equation}
This is a particularly transparent Mars based estimate because it maps a bound on any distributed Hooke-like density into a bound on $K_0$. It agrees very well with the perihelion precession estimate, making our claim more powerful that Mars gives a clean but still weak constraint of order $10^9$ on $|q_0|$.
\\
\\
A more conservative Earth-Mars route comes from the traditional cosmological constant bound \cite{jetzer2006two,sereno2006solar,iorio2008solar}
\begin{equation}
\Lambda \lesssim 10^{-36} {\rm km}^{-2}
\end{equation}
For a cosmological constant, we have 
\begin{equation}
K_\Lambda = \frac{\Lambda c^2}{3}
\end{equation}
Converting $10^{-36} {\rm km}^{-2} = 10^{-42} {\rm m}^{-2}$ one finds
\begin{equation}
K_\Lambda \lesssim \frac{(3.0 \times 10^8)^2}{3}10^{-42} {\rm s}^{-2}
\end{equation}
this means that
\begin{equation}
K_\Lambda \lesssim 3.0 \times 10^{-26} {\rm s}^{-2}
\end{equation}
which can also be written as 
\begin{equation}
K_\Lambda \lesssim 3.0 \times 10^{-11} {\rm yr}^{-2}
\end{equation}
The corresponding deceleration parameter estimate is
\begin{equation}
|q_0| \lesssim \frac{3.0 \times 10^{-11}}{4.75 \times 10^{-21}} |q_0| \lesssim 6.3 \times 10^9
\end{equation}
This is, again, weaker than the modern Mars perihelion or dark density mapping estimates, but it is nevertheless useful as a conservative benchmark showing that solar system measurements are many orders of magnitude away from testing a cosmological value $q_0 \sim -1$. \\ 

\subsection{Limits from the outer solar system route}
The outer solar system is quite attractive for our prospects because the anomalous acceleration $A_K = K_0r$ grows with distance. However, the precision and cleanliness of orbital modeling becomes much worse for trans Neptunian objects, long period comets and Oort-cloud bodies. \cite{brown2004discovery,vokrouhlicky2019origin} So, what is key here is one must distinguish between precision ephemeris bounds and rough dynamical consistency bounds. 
For a detached trans Neptunian object such as Sedna, one can again begin with the perihelion precession estimate \cite{brown2004discovery}
\begin{equation}
|K_0| \lesssim \frac{2n|\Delta \dot\varpi|}{3\sqrt{1-e^2}}
\end{equation}
Taking representative Sedna like parameters $a_{\rm Sedna} \simeq 541.6 {\rm AU}$ and $e_{\rm Sedna} \simeq 0.859$, Kepler's law gives
\begin{equation}
P_{\rm Sedna} \simeq a_{\rm Sedna}^{3/2} {\rm yr}
\end{equation}
so that
\begin{equation}
P_{\rm Sedna} \simeq 1.26 \times 10^4 {\rm yr}
\end{equation}
and
\begin{equation}
n_{\rm Sedna} = \frac{2\pi}{P_{\rm Sedna}} \simeq 4.99 \times 10^{-4} {\rm yr}^{-1}
\end{equation}
If one optimistically assumes $\Delta \dot\varpi_{\rm Sedna} \sim 1 {\rm arcsec \ cty}^{-1}$ then $1 {\rm arcsec \ cty}^{-1} = 4.848 \times 10^{-8} {\rm rad \ yr}^{-1}$ and this substitution gives us
\begin{equation}
|K_0|_{\rm Sedna} \lesssim 3.1 \times 10^{-11} {\rm yr}^{-2}
\end{equation}
and therefore
\begin{equation}
|q_0|_{\rm Sedna} \lesssim 6.6 \times 10^9
\end{equation}
If instead the effective anomalous precession sensitivity were improved to
\begin{equation}
\Delta \dot\varpi_{\rm Sedna} \sim 0.1 {\rm arcsec \ cty}^{-1}
\end{equation}
then the limit would scale linearly as
\begin{equation}
|q_0|_{\rm Sedna} \lesssim 6.6 \times 10^8
\end{equation}
This shows why distant objects are potentially attractive for our pursuit here. The reason is that their small mean motion enhances the precession response. The difficulty meanwhile is that present TNO data cover only a small fraction of such long period orbits, so a clean anomalous precession measurement is not realistically available at this level. A more direct short arc estimate can be obtained by treating the anomalous acceleration as producing a cumulative angular residual. Dimensionally, we have $\delta \theta \sim \frac{1}{2}|K_0|T^2$ so that
\begin{equation}
|K_0| \lesssim \frac{2\delta \theta}{T^2}
\end{equation}
and hence we have
\begin{equation}
|q_0| \lesssim \frac{2\delta \theta}{H_0^2T^2}
\end{equation}
For a Sedna like observed arc $T \simeq 35.4 {\rm yr}$ and an astrometric residual scale $\delta \theta = 0.1 {\rm arcsec} = 4.848 \times 10^{-7} {\rm rad}$ one obtains
\begin{equation}
|q_0| \lesssim \frac{2(4.848 \times 10^{-7})}{(4.75 \times 10^{-21})(35.4)^2} \lesssim 1.6 \times 10^{11}
\end{equation}
If instead one assumes milliarcsecond level effective astrometry, $\delta \theta = 1 {\rm mas} = 4.848 \times 10^{-9} {\rm rad}$
then
\begin{equation}
|q_0| \lesssim 1.6 \times 10^9
\end{equation}
This short arc estimate is less favorable than the ideal perihelion precession estimate. It also is interesting in the sense that it captures the practical obstacle for TNOs, which is that the orbital baseline is short compared with the orbital period, so the fitted orbit can absorb much of the anomalous acceleration.
\\
\\
For long period comets, the same short arc argument applies \cite{vokrouhlicky2019origin,womack2021visual} and taking a Hale Bopp-like observational arc $T \simeq 29.2 {\rm yr}$
and using that we have $\delta \theta = 0.1 {\rm arcsec}$ and that
gives us
\begin{equation}
|q_0| \lesssim \frac{2(4.848 \times 10^{-7})}{(4.75 \times 10^{-21})(29.2)^2} 2.4 \times 10^{11}
\end{equation}
If the residual scale were instead
\begin{equation}
\delta \theta = 0.01 {\rm arcsec}
\end{equation}
then
\begin{equation}
|q_0| \lesssim 2.4 \times 10^{10}
\end{equation}
long period comets are therefore not competitive as precision probes of $q_0$. It is so because their large semimajor axes are not enough to overcome the short observed arcs, nongravitational outgassing forces, and strong perturbations from the Galactic tide and passing stars. \cite{vokrouhlicky2019origin,torres2019galactic}
\\
\\
The Oort-cloud route is different and that reason behind that is that it is not a precision orbit measurement but a dynamical consistency argument. \cite{brasser2013oort,iorio2012impact} Requiring the cosmological acceleration to be smaller than solar gravity at a radius $R$ gives
\begin{equation}
|q_0|H_0^2R \lesssim \frac{GM_\odot}{R^2}
\end{equation}
or
\begin{equation}
|q_0| \lesssim \frac{GM_\odot}{H_0^2R^3}
\end{equation}
Using
\begin{equation}
GM_\odot = 4\pi^2 {\rm AU}^3{\rm yr}^{-2}
\end{equation}
one obtains
\begin{equation}
|q_0| \lesssim \frac{4\pi^2}{H_0^2R^3}
\end{equation}
where $R$ is measured in AU and so for $R = 5 \times 10^4 {\rm AU}$
this gives
\begin{equation}
|q_0| \lesssim 6.7 \times 10^7
\end{equation}
While for $R = 10^5 {\rm AU}$ one obtains
\begin{equation}
|q_0| \lesssim 8.3 \times 10^6
\end{equation}
and for $R = 2 \times 10^5 {\rm AU}$ one obtains
\begin{equation}
|q_0| \lesssim 1.0 \times 10^6
\end{equation}
On the surface, these numbers are stronger than the clean Mars and Saturn ephemeris estimates. But they depend on interpreting the existence and stability of the distant Oort cloud as a constraint on an additional Hooke-like acceleration, which is less direct than a planetary ephemeris fit. Another Oort-cloud estimate can be made by requiring that the cosmological acceleration can not exceed a characteristic Galactic tidal acceleration. \cite{klacka2010galactic,heisler,torres2019galactic} With that in mind, we write
\begin{equation}
|q_0|H_0^2R \lesssim A_{\rm tide}
\end{equation}
so that
\begin{equation}
|q_0| \lesssim \frac{A_{\rm tide}}{H_0^2R}
\end{equation}
Taking $A_{\rm tide} \sim 10^{-14} {\rm m \ s}^{-2}$ and using $H_0$ in SI units gives $H_0^2 = 4.75 \times 10^{-36} {\rm s}^{-2}$. This means that for $R = 2 \times 10^4 {\rm AU}$ one finds
\begin{equation}
|q_0| \lesssim 7.0 \times 10^5
\end{equation}
While for $xR = 5 \times 10^4 {\rm AU}$ one finds
\begin{equation}
|q_0| \lesssim 2.8 \times 10^5
\end{equation}
and for $R = 10^5 {\rm AU}$
one finds
\begin{equation}
|q_0| \lesssim 1.4 \times 10^5
\end{equation}
These are the strongest numerical estimates among all the routes  we have considered so. Their big weakness is not numerical but rather a bit conceptual, which is that the Galactic tide is itself part of the physics governing Oort-cloud dynamics. So, the inferred population is subject to large uncertainties, stellar encounters, and selection effects. \cite{brasser2013oort,klacka2010galactic,torres2019galactic}
\\
\\
Collecting the estimates, the clean ephemeris level constraints are of order
\begin{equation}
|q_0|_{\rm Mars/Saturn} \sim 10^9
\end{equation}
while the TNO and long period comet direct orbit constraints are roughly
\begin{equation}
|q_0|_{\rm TNO/LPC} \sim 10^9{\rm \ to \ }10^{12}
\end{equation}
and the Oort cloud dynamical consistency estimates are roughly
\begin{equation}
|q_0|_{\rm Oort} \sim 10^5{\rm \ to \ }10^8
\end{equation}
Thus, the cleanest bounds are still Mars and Saturn at the level of a few times $10^8$ to $10^9$, whereas the outermost solar system can give stronger but more model-dependent dynamical bounds.  
\\

\section{Statistical Uncertainty Analysis}
We now move forward to statistical analysis to systematically account for the uncertainties and assess the robustness of the resulting bounds. First, for the perihelion based constraints, we propagate the uncertainties in the measured orbital parameters through $K_0$ and $q_0$ using Monte Carlo analysis. The Monte Carlo analysis gives a median $|q_0| = 5.75 \times 10^8$ for Saturn with a $95\%$ upper limit of $1.60 \times 10^9$ while for Mars the corresponding values are $6.59 \times 10^8$ and $1.88 \times 10^9$. So after propagating the uncertainties, the clean planetary constraints still remain at roughly the $10^9$ level  as in our original analysis. Monte Carlo analysis for the local dark matter density route gives a constraint of roughly $|q_0| < 8.2 \times 10^8$ which is close to the Mars perihelion result and the cosmological constant benchmark gives a weaker constraint of $|q_0| < 6.3 \times 10^9$ but still provides a useful comparison. Sedna's formal perihelion based constraint gives a median $|q_0| = 6.44 \times 10^9$ and a $95\%$ upper limit of $6.89 \times 10^9$. However, this is clearly the more idealized case, since only a small fraction of Sedna's very long orbit is observed which is why we are treating the outer Solar System constraints more cautiously.

\begin{table}[ht]
\centering
\caption{Monte Carlo constraints on the magnitude of the deceleration
parameter, $|q_0|$. The median and 95\% upper limit values are obtained
from the corresponding Monte Carlo distributions.}
\label{tab:q0_constraints}
\begin{tabular}{lcc}
\hline
\textbf{Case} & \textbf{Median $|q_0|$} & \textbf{95\% upper limit} \\
\hline
Saturn perihelion       & $5.75\times10^{8}$  & $1.60\times10^{9}$ \\
Mars perihelion         & $6.59\times10^{8}$  & $1.88\times10^{9}$ \\
Sedna perihelion        & $6.44\times10^{9}$  & $6.89\times10^{9}$ \\
Mars density            & $8.21\times10^{8}$ & $8.79\times10^{8}$ \\
Cosmological constant   & $6.60\times10^{9}$  & $1.67\times10^{10}$ \\
\hline
\end{tabular}
\end{table}
\begin{comment}

\begin{figure}[t]
    \centering
    \includegraphics[width=\columnwidth]{monte_carlo_summary.png}
    \caption{Monte Carlo constraints on \(K_0\) and \(q_0\) obtained by propagating the uncertainties in the orbital and precession parameters. The distributions show the inferred values of \(|K_0|\) and \(|q_0|\), with the median and 95\% upper limits indicated by the dashed and dotted lines, respectively.}
    \label{fig:mc_constraints}
\end{figure}
\end{comment}
To assess how strongly the derived constraints depend on the precision of the perihelion precession measurements, we perform a sensitivity analysis in which the uncertainty in the measured perihelion precession rate $\sigma_{\Delta\dot{\varpi}}$ is varied over the range $10^{-4}$-$1~\mathrm{mas~century}^{-1}$ while keeping the central residual and the orbital parameters fixed. The resulting upper bound (with 95\% confidence) on $|q_0|$ for Saturn, Mars and Sedna is shown in Figure 1(b). As the observational uncertainties decrease, the constraints from Saturn and Mars become progressively tighter, showing that the inferred upper bound on $|q_0|$ is directly dependent on the precision of the orbital measurements, whereas the upper bound for Sedna exhibits comparatively little variation across the explored uncertainty range. The improvement is especially strong for Mars and Saturn when the uncertainties are larger, and gets progressively flatter towards smaller uncertainties, indicating that further reductions in the measurement uncertainty lead to increasingly smaller improvements in the constraint.
\\
\\
For the precisions adopted, the resulting bounds are of order $|q_0| \lesssim 10^9$ for Saturn, Mars and Sedna. This difference emphasizes that the dynamical sensitivity of the orbit and the attainable accuracy of the perihelion precession measurement are both important. The sensitivity analysis shows that this would improve greatly constraints on a possible cosmological contribution to local Solar System dynamics, if it is measured more precisely, especially in the case of distant objects like Sedna.
\\
\\
In Figure 1(c), we show the joint dependence of the $|q_0|$ on the the orbital eccentricity $e$ and the perihelion precession residual $|\Delta \dot\varpi|$ for Saturn, Mars and Sedna. We see that the bound becomes stronger for small values of $|\Delta \dot\varpi|$ while the dependence on eccentricity $e$ is comparatively weak over most of the parameter space. The contours also show that the sensitivity to the perihelion residual becomes more pronounced as eccentricity $e$ approaches unity. The fiducial parameter marked by asterisk lies within the regions consistent with the bounds obtained from our Monte Carlo analysis. The figure therefore provides a broader view of how the inferred constraint changes across the orbital space and confirms that the perihelion precession residual is a significant parameter in controlling the strength of the $|q_0|$ bound with a secondary dependence on eccentricity. The numerical bounds obtained from the fiducial orbital parameters should not be interpreted as universal, model independent point constraints on $q_0$. The inferred bound depends on the measured perihelion precession residual, its observational uncertainty, and the orbital parameters entering the mapping to $K_0$. The planetary constraints remain the same at $\sim10^9$ within the uncertainty propagation, while Sedna provides a much stronger constraint and could be considered as a more idealized outer Solar System constraint given the limited orbital coverage. The statistical analysis gives a description of the dependence of the constraint on $|q_0|$ on the available orbital measurements and uncertainties. 
\begin{figure}[H]
    \centering
    \includegraphics[
        width=0.95\textwidth,
        height=0.72\textheight,
        keepaspectratio
    ]{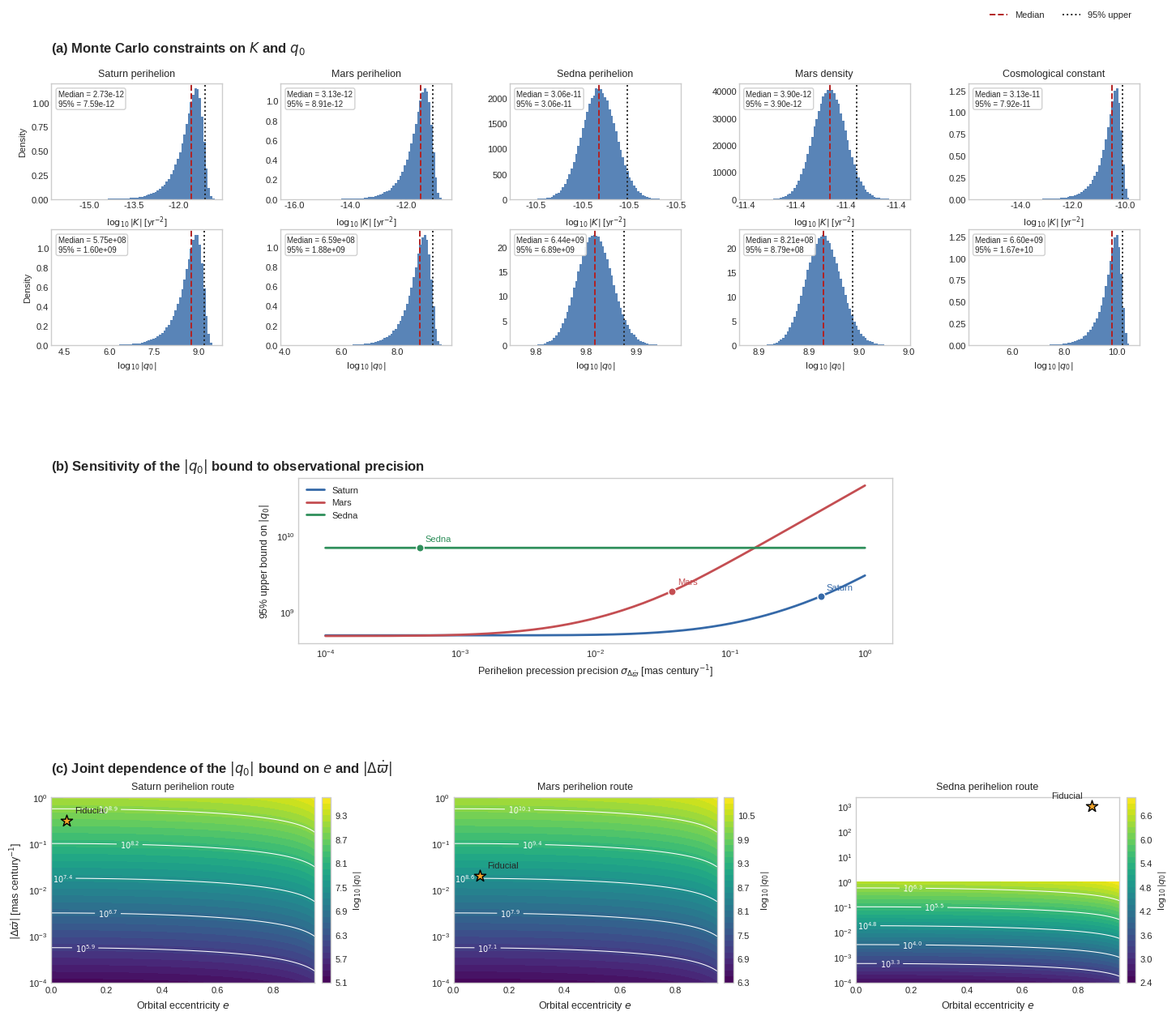}
    \caption{
    Statistical constraints and sensitivity analysis for the cosmological coupling parameter $K_0$ and deceleration parameter $q_0$.
    (a) Monte Carlo distributions of $|K_0|$ and $|q_0|$ for the Saturn perihelion, Mars perihelion, Sedna perihelion, Mars dark matter density, and cosmological constant routes. Red dashed lines indicate the median values, while black dotted lines indicate the 95\% upper bounds.
    (b) Sensitivity of the 95\% upper bound on $|q_0|$ to the assumed perihelion-precession precision for Saturn, Mars, and Sedna.
    (c) Joint dependence of the $|q_0|$ bound on orbital eccentricity $e$ and the adopted precession sensitivity $|\Delta\dot{\varpi}|$ for the Saturn, Mars, and Sedna perihelion routes. Stars indicate the fiducial parameter choices.
    }
    \label{fig:statistical_constraints}
\end{figure}
\FloatBarrier
The results obtained from the statistical analysis form the basis for the scope of constraints that can be currently obtained from Solar System probes, given the available observational precision and orbital parameters. Statistical analysis can be further extended by combining individual constraints within a joint likelihood framework to obtain a unified constraint on $K_0$. This would allow the available observational information from different Solar System probes to be incorporated simultaneously, potentially yielding tighter constraints on $q_0$. In this approach, each measurement can be used to construct an individual likelihood function and, assuming statistical independence, the overall likelihood can be calculated as the product of individual likelihoods.The joint likelihood would combine the available observational information into a single statistical framework, which could then be combined with an appropriate prior distribution using Bayes' theorem to obtain the posterior probability distribution of $K_0$.The posterior distribution could then be converted to a constraint on the cosmological deceleration parameter using the theoretical relationship between $K_0$ and $q_0$. An extension of this and other statistical methods, like the Robert–Fischer method, could give a more holistic view of the constraints and their consistency.
\\
\\
The constraints that we have derived from the small deviations in Solar System and outer Solar System are also prone to local Newtonian perturbations and other astrophysical disturbances. Particularly, the measured orbital residuals might not have been due to any cosmological contribution, the gravitational effect of the planets, the asteroid belt, Kuiper belt objects and other matter are already incorporated to varying degrees in modern planetary ephemerides, but uncertainties in their masses and spatial distributions can contribute to the residual dynamical signal. Asteroid perturbations for example are also a source of systematic uncertainty in the planetary ephemerides, especially for Mars, and special asteroid masses and the omission of perturbers have been developed with the aim of minimizing these effects. \cite{pitjev2013constraints, fienga, kuchynka}. Similarly for the outer planets, the Kuiper belt induced perturbations can be measured and taken into account for small residual accelerations and for looking for other gravitational effects.\cite{pitjeva2018mass, iorio2007} 
\\
\\
Another source of systematic uncertainty in planetary orbital dynamics is the solar quadrupole moment $J_2$. Solar oblateness produces a characteristic perihelion precession whose dependence on orbital distance differs from that of the cosmological Hooke-like term, and therefore it is not directly degenerate with $K_0$. However, the uncertainties in the measured or fitted value of $J_2$ and correlations between the various parameters in global ephemeris solutions may contribute to small residuals in the orbit and hence to the limits on additional accelerations\cite{fienga2011, park}. This effect should be easier to distinguish from a cosmological contribution due to its unique orbital geometry and distance dependence.
\\
\\
The Galactic tidal field also causes a differential acceleration throughout the Solar System. Examples of Galactic contributions have been taken into account in the studies of dynamics in the Solar system and tests of small non-Newtonian or cosmological effects\cite{iorio2006solar,iorio2008solar, sereno2006solar}. The Galactic tidal field is not necessarily isotropic and in linear $r$ form like the cosmological component considered here but the uncertainty of the local Galactic potential, such as the distribution of dark matter in the area near the Solar System, does not allow the isolation of a very small cosmological contribution with high precision \cite{iorio2006solar}. Galactic tides become more significant for very far away objects in the Solar system, especially for the long term evolution of the orbit of the members of the Oort cloud \cite{heisler}. These effects do not dominate the dynamics of Sedna for the distances that we have analyzed here but are still an external perturbation which should be taken into consideration when explaining the progressively smaller orbital residuals as a cosmological signal.
\\
\\
The constraints we have derived are not just confined to Solar System but it also presents what they could mean for cosmological dynamics and whether similar effects might be looked for in other gravitationally bound astrophysical systems. It is still an interesting theoretical question whether such systems are independent of the expansion of the universe or whether they get a suppressed local imprint of the cosmological background. The interiors of gravitationally bound systems are, for instance, not sensitive to the global expansion as proposed by Wang and Wang\cite{wang2023decoupling} while alternative models have suggested cosmological coupling via compact object evolution. This could be tested using related effects in other astrophysical systems. Stellar binaries and binary pulsars are complementary probes via precession and timing, respectively, while galaxy groups and clusters are probes of the competition between gravitational attraction and cosmic acceleration via their turnaround radii.
\\
\\
For the sake of completeness, it is also worthwhile to briefly compare the present observational determinations of the cosmological
deceleration parameter alongwith the limits obtained from Solar System dynamics here. A variety of cosmographic reconstructions employing Type Ia supernovae, cosmic chronometers, baryon acoustic oscillations and fast radio bursts have obtained estimates of $q_0$ typically of order unity, summarized in Fig. \ref{qo}, with many analyses favouring negative values corresponding to accelerated expansion \cite{q01cattoen2007cosmography,q02hu2025constraints,q03koussour2026thermodynamically,q04son2025strong,q05kr2026unified,q06verdugo2026reconstructing,q07sales2026cosmographic,q08lemos2026impact}. Still, the inferred values depend on the observational datasets, adopted parameterizations and treatment of astrophysical systematic uncertainties.
\clearpage

\begin{figure}[!ht]
    \centering
    \includegraphics[width=1\linewidth]{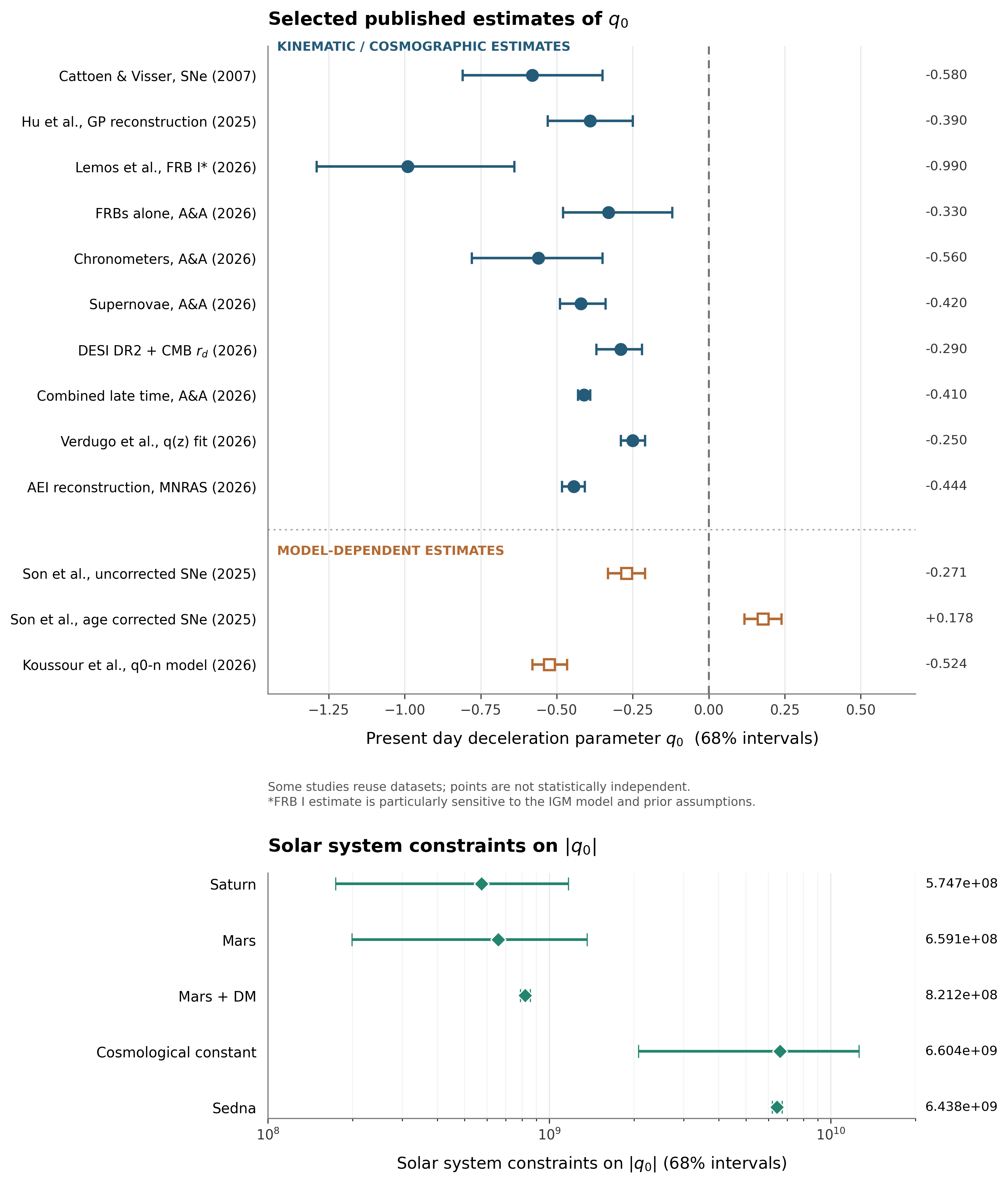}
    \caption{Whisker plot showing selected literature estimates of the present day cosmological deceleration parameter $q_0$, with reported $1\sigma$ uncertainty intervals and our derived Solar System constraints. The displayed estimates are subject to different modeling assumptions and some share observational dataset.}
    \label{qo}
\end{figure}

\section{Future Improvements}
One natural extension of Solar System probes is on other stellar systems in which orbital dynamics can be measured independently. In works of Jetzer and Sereno\cite{jetzer2006two}, they discussed constraints from planetary observations and binary pulsars and looked at the effect of the cosmological constant on the two body problem. They also considered stellar system observables including periastron advance and gravitational redshift as possible probes of the cosmological constant \cite{sereno2006solar}. These systems are very valuable alongside being the basis for independent orbital configurations they can also be used to test predictions of other dynamical contributions. More specifically, binary pulsars allow for accurate timing measurements of the orbit and for the precise measurement of relativistic effects, but an intrinsic contribution to the orbit evolution and kinematic effects must also be taken into account for the extraction of a cosmological contribution. A comparison with the present work is thus methodological: both methods aim to limit small deviations from the expected dynamics of a bound system, with the only difference being that the bounds need not be for the same physical parameter.
\\
\\
On galactic and larger scales, it is possible to consider a different approach, where the local gravitational pull and cosmological acceleration become comparable and can affect the outer edges of bound structures. A characteristic zero acceleration radius is determined by balancing the gravitational attraction and repulsive term of the cosmological constant in a spherical mass distribution in a universe containing a cosmological constant. The turnaround radius argument gives an upper bound on the bound structure size for a given mass within the standard $\Lambda$CDM cosmology for a given argument.  Numerical simulations of the galaxy clusters have been used to test these predictions and it is found that turnaround scales can be detected even in non-spherical systems, albeit with a dependence on tidal effects from neighbouring massive systems. In recent years, the treatments of zero acceleration and turnaround surfaces have been extended to galaxy groups with general mass distributions. These systems also provide alternative methods for exploring the interactions between bound systems and the cosmological background. 
\\
\\
Cosmological coupling has also been discussed in a different context, the possible evolution of black hole masses with the scale factor. 
\\
\\
What is the most interesting thing with these constraints is that these have quite some possibilities of improvements. The most straightforward way to improve these constraints is to fit $K_0$ directly in modern or future planetary ephemerides. A post fit translation from perihelion residuals is useful for order of magnitude estimates, but a dedicated fit would account for correlations with $GM_\odot$, asteroid masses, solar oblateness, initial conditions, $\dot G/G$ and other small accelerations. For perihelion based constraints, the scaling is simple, which is that
\begin{equation}
|q_0| \propto n\Delta \dot\varpi
\end{equation}
Thus, a factor $\mathcal{I}$ improvement in anomalous precession sensitivity produces
\begin{equation}
|q_0|_{\rm new} \simeq \frac{|q_0|_{\rm old}}{\mathcal{I}}
\end{equation}
Starting from a Mars or Saturn level of $|q_0|_{\rm old} \sim 10^9$,
a tenfold improvement would give
\begin{equation}
|q_0|_{\rm new} \sim 10^8
\end{equation}
and a hundredfold improvement would give
\begin{equation}
|q_0|_{\rm new} \sim 10^7
\end{equation}
This is the realistic target range for improved Mars ranging, Cassini era reanalysis, or interplanetary laser ranging. Even such improvements would still be far from testing $q_0$ of order unity, though.
\\
\\
Future ice-giant orbiters could be more powerful because the precession sensitivity improves for small $n$, as by comparing an outer planet $p$ with Saturn, one has
\begin{equation}
\frac{|q_0|_p}{|q_0|_{\rm Sat}} \simeq \frac{n_p}{n_{\rm Sat}}\frac{\Delta \dot\varpi_p}{\Delta \dot\varpi_{\rm Sat}}
\end{equation}
For Uranus, $P_{\rm Uranus} \simeq 84 {\rm yr}$ so that
\begin{equation}
\frac{n_{\rm Uranus}}{n_{\rm Sat}} \simeq \frac{P_{\rm Sat}}{P_{\rm Uranus}} \simeq 0.35
\end{equation}
For Neptune, $P_{\rm Neptune} \simeq 165 {\rm yr}$ so that
\begin{equation}
\frac{n_{\rm Neptune}}{n_{\rm Sat}} \simeq \frac{P_{\rm Sat}}{P_{\rm Neptune}} \simeq 0.18
\end{equation}
If a future Uranus orbiter reached $\Delta \dot\varpi_{\rm Uranus} \sim 0.1 {\rm mas \ cty}^{-1}$ then the expected sensitivity would be approximately $|q_0|_{\rm Uranus} \sim 5 \times  10^7$, while if it reached $\Delta \dot\varpi_{\rm Uranus} \sim 0.01 {\rm mas \ cty}^{-1}$ then
\begin{equation}
|q_0|_{\rm Uranus} \sim 5 \times 10^6
\end{equation}
Similarly, for a Neptune orbiter one could plausibly reach
\begin{equation}
|q_0|_{\rm Neptune} \sim 3 \times 10^7
\end{equation}
for a $0.1 {\rm mas \ cty}^{-1}$ sensitivity and
\begin{equation}
|q_0|_{\rm Neptune} \sim 3 \times 10^6
\end{equation}
for a $0.01 {\rm mas \ cty}^{-1}$ sensitivity. These are among the best clean future solar system possibilities, because they combine large orbital scale with spacecraft-level tracking.
\\
\\
For TNOs, the relevant improvement comes mainly from longer baselines and better astrometry.The short arc scaling gives $|q_0| \lesssim \frac{2\delta \theta}{H_0^2T^2}$ so that
\begin{equation}
|q_0|_{\rm new} \simeq |q_0|_{\rm old}\left(\frac{\delta \theta_{\rm new}}{\delta \theta_{\rm old}}\right)\left(\frac{T_{\rm old}}{T_{\rm new}}\right)^2
\end{equation}
If a Sedna like object improves from $\delta \theta_{\rm old} = 0.1 {\rm arcsec}$ to $\delta \theta_{\rm new} = 1 {\rm mas}$
and the baseline improves from $T_{\rm old} = 35 {\rm yr}$
to $T_{\rm new} = 70 {\rm yr}$ then the improvement factor is
\begin{equation}
\left(\frac{1 {\rm mas}}{0.1 {\rm arcsec}}\right)\left(\frac{35}{70}\right)^2 = 2.5 \times 10^{-3}
\end{equation}
Thus a present estimate of $|q_0|_{\rm old} \sim 10^{11}$
could become $|q_0|_{\rm new} \sim 3 \times 10^8.$ This goes to show that TNOs can become competitive over long time baselines, but probably not immediately. Their power grows substantially only when milliarcsecond astrometry and multi-decade arcs are combined. For the Oort cloud, future improvements would not come from precision orbit fitting of individual objects but from population level modeling. If the effective radius at which the cloud is dynamically confirmed increases from $R_1$ to $R_2$, the solar binding bound scales as
\begin{equation}
|q_0| \propto R^{-3}
\end{equation}
so that
\begin{equation}
\frac{|q_0|_2}{|q_0|_1} = \left(\frac{R_1}{R_2}\right)^3
\end{equation}
For example, extending a robust dynamical inference from $R_1 = 5 \times 10^4 {\rm AU}$ to $R_2 = 10^5 {\rm AU}$ would improve the binding estimate by
\begin{equation}
\left(\frac{5 \times 10^4}{10^5}\right)^3 = \frac{1}{8}
\end{equation}
The Galactic tide comparison instead scales as
\begin{equation}
|q_0| \propto R^{-1}
\end{equation}
so increasing the effective radius by a factor of two improves that estimate only by a factor of two. Hence the Oort cloud route can remain numerically powerful, but its uncertainty will remain dominated by astrophysical modeling rather than measurement precision alone. Overall, next generation measurements could plausibly shift the clean solar system constraints from $|q_0| \sim 10^9$ to $|q_0| \sim 10^7{\rm \ to \ }10^8$  with improved Mars and Saturn analyses and perhaps to $|q_0| \sim 10^6{\rm \ to \ }10^7$ with a dedicated Uranus or Neptune orbiter. TNO and Oort-cloud routes could become useful complementary probes, but their strongest numerical limits would require careful population level modeling and it would be less clean than planetary ephemeris constraints.
\\

\section{Conclusions}
In this work we have investigated the possibility of constraining the present day cosmological deceleration parameter through Solar System dynamics, hence exploring an independent and complementary approach to conventional cosmological observations. By relating the cosmic acceleration to a weak Hooke type radial perturbation, we have developed a framework for translating anomalous orbital effects into constraints on the deceleration parameter. We examined several observational routes, including planetary perihelion precession measurements for Mars and Saturn, the orbital dynamics of trans-Neptunian objects and long period comets and the dynamical stability of the Oort cloud. While the planetary observations provide comparatively robust constraints, the outer Solar System offers potentially greater sensitivity at the expense of increased dependence on astrophysical assumptions and dynamical modeling. We have further investigated the statistical robustness of these estimates through Monte Carlo uncertainty propagation and sensitivity analyses, alongside a discussion of systematic uncertainties arising from local gravitational perturbations. Our results demonstrate that although present Solar System observations remain far from achieving the sensitivity required to distinguish between cosmologically relevant values of the deceleration parameter, they provide a useful framework for examining possible imprints of cosmological acceleration on gravitationally bound systems.
\\
\\
The broader implications of this work lie in establishing a connection between local gravitational dynamics and the large scale expansion history of the Universe, particularly at a time when evolving dark energy models and possible departures from the standard cosmological picture are receiving increasing observational attention. Future improvements in planetary ephemerides, precision astrometry, interplanetary laser ranging etc. with dedicated missions to the outer planets could progressively strengthen the sensitivity of such independent dynamical tests. Furthermore, extending similar analyses to other gravitationally bound systems, including stellar binaries, binary pulsars and galaxy groups, could provide additional opportunities to investigate the extent to which cosmological evolution leaves observable signatures within local gravitational environments. An important long term objective would be to distinguish genuine cosmological contributions from conventional astrophysical perturbations through increasingly precise and self consistent dynamical models. Although Solar System measurements are unlikely to rival cosmological surveys in constraining the deceleration parameter in the foreseeable future, the framework developed here offers a complementary perspective on the relationship between local gravitational physics and cosmic expansion. In this sense, our investigation contributes to the broader effort to develop independent observational tests of cosmic acceleration and to understand how the dynamical behavior of gravitationally bound systems is connected to the evolution of the Universe.
\section*{Acknowledgements}
The work of O.T. is supported in part by the Vanderbilt Discovery Doctoral Fellowship. The research of M. K. was carried out in the Southern Federal University with financial support from the Ministry of Science and Higher Education of the Russian Federation (State contract FENW-2026-0028). The work of J.L.S. is supported in part by “Xjenza Malta”, as part of the “Technology Development Programme” DTP-2024-014 (CosmicLearning) Project. 

\bibliographystyle{unsrt}
\bibliography{references}

\end{document}